\documentclass{article}
\usepackage[utf8]{inputenc}
\usepackage[english]{babel}
\usepackage[T1]{fontenc}
\usepackage{tabu}
\usepackage{booktabs} 
\usepackage{graphicx}
\usepackage{amssymb}
\usepackage{mathrsfs} 
\usepackage{amsmath} 
\usepackage{authblk} 
\RequirePackage[labelfont=bf]{caption} 

\title{Piketty's second fundamental law of capitalism as an emergent property in a kinetic wealth-exchange model of economic growth}
\author[1]{Quevedo, D. S.}
\author[2]{Quimbay, C. J.}
\affil[1,2]{Department of Physics, Universidad Nacional de Colombia\\Bogot\'a, D.C., Colombia}
\date{\today}

\begin{document}

\maketitle

\begin{abstract}
Emergence of distribution and inequality patterns in closed economic systems have been extensively studied in Econophysics literature by using kinetic models of wealth and income exchange. A basic theoretical approach behind certain type of kinetic models is to consider closed economic systems like ideal gases in such a way that economic transactions with money conservation between pairs of agents are assumed to be analogous to elastic collisions with kinetic energy conservation between pairs of particles. This approach has been the object of numerous criticisms from economists who argue that conservation hypothesis are far from economic reality and constitutes nonsense assumptions. Taking into account these criticisms, as an extension of the most popular kinetic money-exchange model in Econophysics, we propose in this work a kinetic wealth-exchange model of economic growth by introducing saving as a non consumed fraction of production. In this new model, which starts also from microeconomic arguments, it is found that economic transactions between pairs of agents leads the system to a macroscopic behavior where total wealth is not conserved and it is possible to have an economic growth which is assumed as the increasing of total production in time. This last macroeconomic result, that we find both numerically through a Monte Carlo based simulation method and analytically in the framework of a mean field approximation, corresponds to the economic growth scenario described by the well known Solow model developed in the economic neoclassical theory. If additionally to the income related with production due to return on individual capital, it is also included the individual labor income in the model, then the Thomas Piketty's second fundamental law of capitalism is found as a emergent property of the system. We consider that the results obtained in this paper shows how Econophysics can help to understand the connection between macroeconomics and microeconomics.

\end{abstract}

\section{Introduction}
Since the late XXth century, econophysicists had put their efforts towards the study of properties and dynamics involved in financial and economic phenomena. Its contributions in the study of financial markets have been recognized by economists in various ways \cite{Buchanan.2013}. Nevertheless, those recognitions avoid developments in the field of macroeconomics, where wealth and income distributions are studied from empirical and theoretical points of view \cite{Chakraborti.2000,Dragulescu.2000,Dragulescu.2001a,Dragulescu.2001b,Bouchaud.2000,Patriarca.2004a,Saif.2007,Chatterjee.2005,Chakrabarti.2013}. Some economists show skeptical about the validity of theoretical models proposed in this frame \cite{Lux.2005,Galegatti.2006,McCauley.2006}, known as Kinetic exchange models of markets (KEM), where monetary exchanges are modeled inspired in thermodynamic systems in an analogous way of energy exchanges which occur via elastic collisions between pairs of particles of an ideal gas, and distributions are described as emergent properties of microscopic interactions between agents of an economic systems that are matched to simulated and empirical data using mixtures of power laws and generalized exponential distributions.

According to economists the main failure of the KEM is the assumption of conservation of total wealth and income over time, analogous to conservation of total energy, which does not offer any close representation of economic reality \cite{Galegatti.2006}, and moreover, that constitutes a ``\textit{silly}'' assumption from the point of view of modern economics \cite{McCauley.2006}. In line with the last arguments, we propose in this work a kinetic wealth-exchange model of economic growth by introducing saving as a non consumed fraction of production, being it a non-conservative extension of the most popular kinetic money-exchange model developed in Econophysics by Chakraborti and Chakrabarti twenty years ago \cite{Chakraborti.2000}. As a consequence of this new model, we study the emergent properties which build a macroeconomic scenario coherent with Solow model \cite{Solow.1956}, a neoclassical model of economic growth, where Thomas Piketty's second fundamental law of capitalism \cite{Piketty.2014, Piketty.2015, Piketty-Saez.2014, Piketty-Zucman.2013} rises as a macroscopic property of the system.

The model of Chakraborti and Chakrabarti (CC model) \cite{Chakraborti.2000} constitutes a noteworthy approach in the context of KEM which propose an extension of Dragulescu-Yakovenko Model \cite{Dragulescu.2000} (DY model), a model of money distribution in the same spirit of a pure kinetic exchange process between pairs of particles of an ideal gas, by introducing a saving propensity parameter $\lambda$ that limits the available amount of money for each agent at any transaction. For the case of fixed values of $\lambda$, emergent distributions from the model are studied using a gamma-like distribution, a generalized form of Boltzmann's exponential distribution emergent from the DY model \cite{Patriarca.2004b}. In addition, power laws are obtained from the microscopic dynamic of the model constrained to random values of $\lambda$ uniformly distributed over the domain $[0,1]$ \cite{Chatterjee.2004}. The transactions between two economic agents $i$ and $j$ are described by the equations: $m_i(t+1)=m_i(t) + \Delta m$ and $m_j(t+1)=m_j(t) - \Delta m$, where $m_{i,j}(t)$ and $m_{i,j}(t+1)$ are the money that each agent possesses before and after trading, and $\Delta m$ is the amount of money exchanged, which is defined as a function of $m_{i,j}(t)$.

It is easy to note from the last equations, that for every point of time it is satisfied that $m_i(t)+m_j(t)=m_i(t+1)+m_j(t+1)$, thus the total money of the system remains constant through time. As we emphasized before, this fact constitutes one of the main subjects of criticism made by economists. According to Gallegati \textit{et al.} \cite{Galegatti.2006}, this issue arises from the confusion between the concepts of transaction and income within KEM. Properly, the first one constitutes a ``key economic process'' which must be necessarily conservative, and the second is related to production which drives economic growth by ``net physical surplus'', therefore non-conservation is an inherit feature of it. It is followed from the last argument that a more realistic approach to economics should take account of production and not only restricts itself to monetary exchange. This fact was considered by Chakraborti \textit{et al.} \cite{Chakrabarti.2013,Chakrabarti.2009} and introduced into a microeconomic framework of the CC model by defining the monetary exchange as a function of the prices of different goods produced and traded by each pair of agents involved in a transaction. Cobb-Douglas utility functions are defined in the CC model, following the spirit of preference model proposed by Silver \textit{et al.} \cite{Silver.2002} to introduce a level of economic rationality to the exchange mechanism aligned with criticisms made by Lux \cite{Lux.2005}, in such a way that the parameter $\lambda$ takes the roll of the weight of money that each agent possesses before trading. Nevertheless, production only offers a mechanism for monetary exchange because goods are assumed to be perishable and no fraction of them is kept after trading. Thus the model remains the spirit of a conservative process.

The last argument gives us a starting point for the extension of the CC model proposed in our work. We consider that saving is not correctly introduced into the CC model, because on the economic sense it is related to keeping a fraction of production which inherently leads to capitalization and economic growth. Thus, we redefine Cobb-Douglas functions from the microeconomic framework of the CC model by introducing a saving parameter $s$ as a non consumed fraction of production. An important characteristic of the redefined functions is that $\lambda$ still remains as a defining parameter, nevertheless it takes the role of the exchange aversion of the economic agents which is exclusively related to the exchange process, and consequently, we name it for the purpose of this paper as exchange aversion parameter. The saving parameter $s$ leads the model to a non-conservative scenario where total wealth and total production increase in time. In this frame, the economic exchange is defined over wealth, because agents save a fraction of production at any trade, conversely to CC model where transactions refer to mere monetary exchange.

In absence of population growth, we measure the economic growth by means of the rate of production growth \cite{Piketty-Zucman.2013}. Using this fact we verify through Monte Carlo simulations that the total production of the system ($Y(t)$) behaves as an exponential function of time-steps ($t$): $Y(t)=Y_0 \exp(gt)$, where $g$ is the economic growth and $Y_0$ is the value of production at $t=0$. A theoretical approach to this result is proposed using a mean field approximation in the same way of Bouchaud and Mezard model of wealth condensation \cite{Bouchaud.2000}. Thus, we investigate the mechanisms for the emergence of exponential growth and we estimate too an analytical expression to predict the value of $g$ as a function of exogenous parameters $s$ and $\lambda$. The consequences of economic growth are studied from the point of view of neoclassical theory of the Solow Model \cite{Solow.1956}, which proposes as a golden rule that on the long run the ratio between wealth and income ($W(t)/Y(t)$) tends to the ratio between the rate of save and rate of economic growth $s/g$. This result in the frame of Piketty's works on economic inequality is called second law of capitalism and it is stated as an empirical regularity for different countries \cite{Piketty.2014,Piketty-Zucman.2013}.

The structure of this paper is proposed as follows: In the second section, we introduce the kinetic wealth-exchange model of economic growth by considering saving $s$ into the same microeconomic framework based on Lagrange multipliers method presented by Chakraborti \textit{et al.} \cite{Chakrabarti.2009}. The macroscopic implications of the kinetic wealth-exchange model is presented in the third section, where we show that wealth distributions are fitted using Log-Pearson distributions, conversely to the CC model. In the fourth section, we present how the economic growth in the context of the Solow model is an emergent macroscopic property of the system using both a Monte Carlo based simulation method and an analytical approximation based in the mean field theory. Finally, in the fifth section, we show as the Piketty's second fundamental Law of capitalism is an emergent property of the system. Conclusions and remarkable ideas are presented in section six.

\section{Kinetic wealth-exchange model of economic \\ growth}
We consider in agreement with DY and CC models  \cite{Dragulescu.2000, Chakraborti.2000} an economic system with a fixed number of agents N which interact pairwise by exchanging a fraction of their wealth in a market (properly in DY and CC models agents exchange money). The economic system is not closed in the sense of conservation of total wealth in time because of the way the model introduces savings; i.e., as a non consumed fraction $s$ of production which increases individual and total wealth of the system. Wealth exchanges occur following the dynamic of a kinetic energy exchange between particles of a thermodynamic system. Thus, two randomly selected agents $i$ and $j$ trade in a market at a time $t$ following the equations: $w_i(t+1)=w_i(t) + \Delta w_i$ and $w_j(t+1)=w_j(t) + \Delta w_j$, where  $w_{i,j}(t)$ and $w_{i,j}(t+1)$ are, respectively, the wealth of each agent before and after trading; and $\Delta w_i$ and $\Delta w_j$ are the rules of interaction between agents. Note that in general $\Delta w_i \neq \Delta w_j$ because they do not restrict themselves to mere exchange, conversely, they consider saving of production, as we will show later.

The exchange mechanisms of the model are defined by introducing production of commodities as it is described below, according to the microeconomic framework of the CC model proposed by Chakrabarti \textit{et al.}  \cite{Chakrabarti.2009,Chakrabarti.2013}, in the spirit of the stochastic model of preferences of Silver \textit{et al}. At any time $t$ before trading, two agents randomly selected from the set of $N$ economic agents produce two different goods. The agent $i$ produces an amount $X$ of one of the commodities, and the agent $j$ produces an amount $Z$ of the other commodity. It is not necessary to specify what kind of commodities are produced, neither that each agent produces the same commodity at any time that it is involved in a trade. However, wealth exchange is defined by the trading prices of commodities $p_{x}$ and $p_{z}$ at time $t$, considering that a fraction $s$ of them is saved after trading, in contrast to CC model, where commodities are assumed to be perishable, i.e. they are completely consumed at every transaction.

The preferences of the economic agents for each commodity are introduced by defining the following utility functions, analogous to the functions introduced into the CC model:
\begin{equation}
U_i(x_i,z_i,w_i(t+1))=\left[(1-s_{x_i})x_i\right]^{\theta_i}\left[(1-s_{z_i})z_i\right]^{\phi_i}w_i^{\lambda_i}(t+1),
\label{eq:QQUi}
\end{equation}
\begin{equation}
U_j(x_j,z_j,w_j(t+1))=\left[(1-s_{x_j})x_j\right]^{\theta_j}\left[(1-s_{z_j})z_j\right]^{\phi_j}w_j^{\lambda_j}(t+1),
\label{eq:QQUj}
\end{equation}
where the terms $(1-s_{x_{i,j}})x_{i,j}$ and $(1-s_{z_{i,j}})z_{i,j}$ are the amounts of each good that each agent consumes which are limited by the rates of saving $s_{x_i}$, $s_{x_j}$, $s_{z_i}$ y $s_{z_j}$. The arguments $x_i$, $x_j$, $z_i$, $z_j$ satisfy the market conditions: $x_i+x_j=X$ and $z_i+z_j=Z$, where $x_i$ and  $z_j$ are the amounts that both agents keep of their own goods in a trading, and $x_j$ and  $z_i$ are the amounts of others good that they buy. The powers on the utility functions $\theta_{i,j}$, $\phi_{i,j}$ are  interpreted as the preferences of agents for each commodity in the same way of the CC model \cite{Chakrabarti.2013,Chakrabarti.2009}; nevertheless, we define the powers of wealth $\lambda_{i,j}$ as the exchange aversion that describes the trend of the economic agents to hold their own wealth at any trading, such that a higher value of $\lambda_{i,j}$ implies a higher resistance to exchange wealth with other agent. For simplicity, we assume in line with Chakrabarti \textit{et al.} \cite{Chakrabarti.2013,Chakrabarti.2009} that the powers on the utility functions are normalized to $1$ as $\theta_{i,j} + \phi_{i,j} + \lambda_{i,j} = 1$.

For the case of zero saving of production, i.e. $s_{x_{i,j}}=s_{z_{i,j}}=0$, both functions reduce to the utility functions introduced into the microeconomic foundation of the CC model \cite{Chakrabarti.2013,Chakrabarti.2009}. The constraints over consumption are introduced using the inequalities show below, which establish that the agents' consumption added to their remaining wealth after trading (left side of the inequality) can not exceed the wealth that they possess before trading, defined as their capital at $t$ $w_{i,j}(t)$ added to the amount of output multiplied by its respective price $p_{x,z}$ (right side of the inequality). Note that the sense of consumption keeps concordance with the utility functions (\ref{eq:QQUi}, \ref{eq:QQUj}), which establish that it is not possible for any agent to consume the whole amount of goods that he buys and holds at any trading because it is limited by the rates of saving $s_{x,z_{i,j}}$. Thus
\begin{equation}
(1-s_{x_i})p_x x_i + (1-s_{z_i})p_z z_i + w_i(t+1) \leq w_i(t) + p_x X,
\label{eq:QQConstrain_i}
\end{equation}
\begin{equation}
(1-s_{x_j})p_x x_j + (1-s_{z_j})p_z z_j + w_j(t+1) \leq w_j(t) + p_z Z.
\label{eq:QQConstrain_j}
\end{equation}
We obtain the rules of interaction between agents $\Delta w_{i,j}$ and the prices of commodities at any transaction from demand functions computed by maximizing the utility functions constrained to equations (\ref{eq:QQConstrain_i}) and (\ref{eq:QQConstrain_j}), following the procedure based on Lagrange multipliers method introduced by Chakraborti \textit{et al.} on the \textit{`Microeconomic foundation of the kinetic exchange models''} \cite{Chakrabarti.2013}. For this purpose we assume that the saving and the exchange aversion of each agent are exogenous parameters which satisfy that $s=s_{x_i}=s_{x_j}=s_{z_i}=s_{z_j}$ and $\lambda=\lambda_i=\lambda_j$.  In this form, the trading prices of commodities are:
\begin{equation}
p_x=\frac{\varepsilon(1-\lambda)[w_i(t)+(w_j(t)]}{X[\lambda-s]},
\label{eq:QQpx}
\end{equation}
\begin{equation}
p_z=\frac{(1-\varepsilon)(1-\lambda)[w_i(t)+(w_j(t)]}{Z[\lambda-s]}.
\label{eq:QQpz}
\end{equation}
The previous results adopt the assumption of the CC model that the preferences of both agents for both goods are equal, i.e. $\theta_i=\theta_j=\theta$ and $\phi_i=\phi_j=\phi$, therefore using the fact that the powers are normalized to $1$ it is defined $\varepsilon=\frac{\theta}{\theta+\phi}$ as a random factor over the domain $[0,1]$.

The expressions for $\Delta w_i$ and $\Delta w_j$ obtained from the trading prices and the demand functions defined from Lagrange multipliers method are shown below. Both functions describe the way in which individual wealth evolve through time, which is a consequence of wealth exchanges occurred between pairs of economic agents and individual saving of production. As we have remarked before $\Delta w_i \neq \Delta w_j$ and total wealth $W(t)$ is not conserved in time. Conversely, $W(t)$ grows as $W(t+1)=W(t)+\Delta w_i + \Delta w_j$, where
\begin{equation}
\Delta w_i = \frac{1-\lambda}{\lambda - s}\left[(s-(1-\varepsilon)\lambda)w_i(t) + \varepsilon\lambda w_j(t) \right],
\label{eq:QQdwi}
\end{equation}
\begin{equation}
\Delta w_j = \frac{1-\lambda}{\lambda - s} \left[(1-\varepsilon)\lambda w_i(t) + (s - \varepsilon\lambda)w_j(t)\right].
\label{eq:QQdwj}
\end{equation}
It is easy to verify from the last equations that for the case of $s=0$ we get that $\Delta w=\Delta w_i=-\Delta w_j$ and the evolution of individual wealth reduces to mere monetary exchange in accordance with the CC model \cite{Chakraborti.2000}. In that scenario the expressions for  $\Delta w_i$ and $\Delta w_j$  reduce to the rule of interaction proposed by Chakraborti and Chakrabarti:  $\Delta m = (1-\lambda)\left[\varepsilon\{ m_i(t)+m_j(t)\}-m_i(t)\right]$ where wealth behaves as simply money ($m$) because there is no saving of production, and commodities are completely consumed at every transaction. The exchange aversion $\lambda$ constitutes the upper limit to $s$ that guarantees the convergence of the model and avoid debt due to $s>\lambda$, thus we restrict the analysis to the cases that satisfy $\lambda > s$.

\section{Macroscopic implications of the model}
\begin{figure}[!ht]
\centering
\includegraphics[scale=0.125]{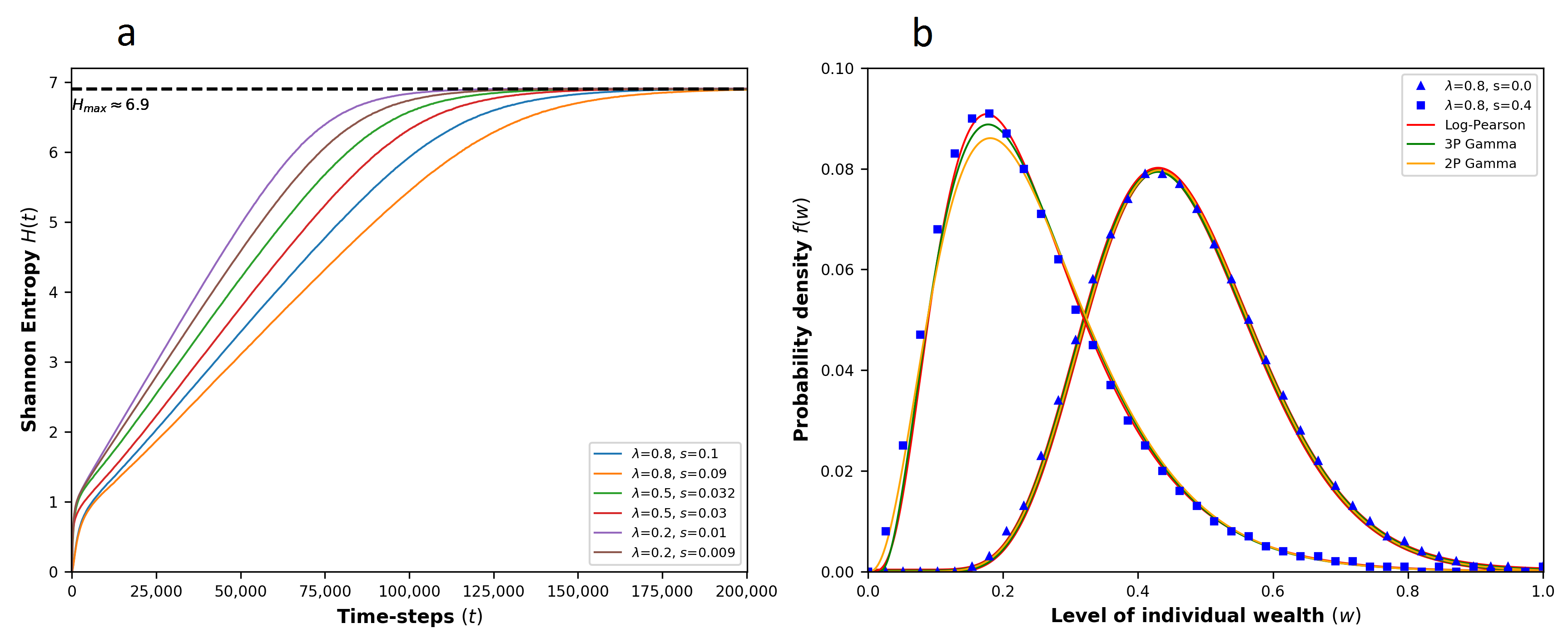}
\caption{\textbf{Steady state and stationary distributions}. (\textbf{a}) The evolution in time of the economic system due to interactions defined by $\Delta w_{i,j}$ is studied through Shannon entropy. The steady state reaches sooner as $s$ become greater or $\lambda$ becomes lower. (\textbf{b}) At this state we found that the wealth distributions are not well-fitted by gamma-like distributions for $s \neq 0$. However all the cases are well-fitted by Log-Pearson distributions.}
\label{Fig:Ht-Dist}
\end{figure}

We study the macroscopic implications of the kinetic wealth-exchange model by using a Monte Carlo based simulation method. Wealth distributions presented in this section, as well as the macroscopic properties discussed in the next one, were obtained for a fixed number of agents $N=1000$ which interact pairwise according to the equations (\ref{eq:QQdwi}) and (\ref{eq:QQdwj}). We assigned for every agent an initial individual wealth $w_i|_{t=0}=1$ such that the total wealth of the system starts at $W|_{t=0}=1000$. The steady state of the system is studied using the Shannon entropy. In the figure \ref{Fig:Ht-Dist}-a we show the pattern of maximization of the entropy for different sets of values of $s$ and $\lambda$. It is clear that the system reaches faster the steady state as $s$ become larger, conversely, the increasing of $\lambda$ slows down the process of relaxation because the exchange aversion of economic agents become larger.

The wealth distributions at equilibrium are obtained over $t=5 \times 10^5 $ time-steps taking an average over $10^{5}$ ensembles. The behavior of the entropy and its dependence of exogenous parameters $\lambda$ and $s$ ensure that at this point of time the steady equilibrium is already reached for every case studied. According to the CC model \cite{Patriarca.2004b}, stationary distributions are well-fitted by two parameter gamma distributions $f(w)=\frac{w^{\gamma-1}}{\delta^{\gamma} \Gamma(\gamma)}\exp\left(-w/\delta\right)$, where $\delta$ and $\gamma$ are the shape and the scale of the distribution that satisfy $\delta \dot{ } \gamma = \langle w \rangle$. That situation constitutes the limit case $s=0$ of the extended model that we introduce. Nevertheless, we found that the goodness of fit decreases as $s \rightarrow \lambda$, because the expressions obtained for $\Delta w_{i,j}$ behaves more volatile as consequence of the term $\frac{1}{\lambda - s}$ and the individual wealth reaches extremely high values. As an alternative to this issue we introduce the Log-Pearson distribution, in the figure \ref{Fig:Ht-Dist}-b we show simulated data for the cases $\lambda=\{0.8\}$, $s=\{ 0.0, 0.4\}$ fitted by three and two parameters gamma distributions, and by Log-Pearson distributions. For $s=0.0$ it is clear that the data is well-fitted by the three distributions, however for $s=0.4$ the data does not match to any gamma distribution.

\begin{figure}[!ht]
\centering
\includegraphics[scale=0.125]{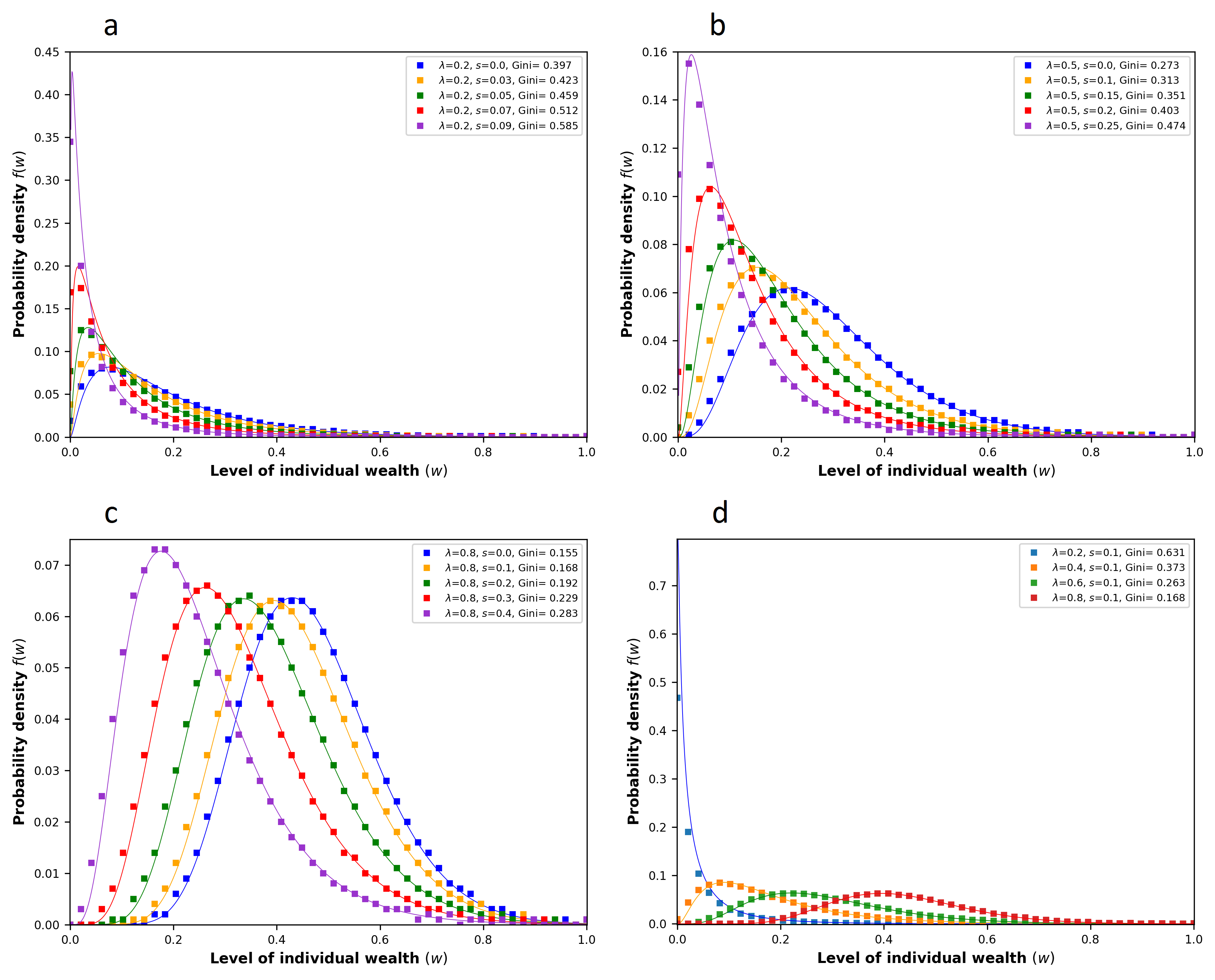}
\caption{\textbf{Emergent wealth distributions.} (\textbf{a-c}) Simulated data for fixed values of  $\lambda= \{0.2$, $0.5$, $0.8\}$ and different values of $s$ is well-fitted by Log-Pearson distributions. The distributions become more equitable as $s$ become larger. (\textbf{d}) For the case of fixed $s=0.1$ the distribution becomes more equitable as the exchange aversion becomes greater, following the behavior of the CC model according. All results are normalized to $1$ to be presented in the same scale, the level of inequality of each distribution is measured through the Gini index shown in each figure.}
\label{Fig:LP}
\end{figure}

The distributions for fixed values of  $\lambda= \{0.2$, $0.5$, $0.8\}$ and different values of $s$  fitted by Log-Pearson distributions are shown in the figures \ref{Fig:LP}-a to \ref{Fig:LP}-c, for all the cases the data has been normalized to 1  to be presented in the same scale dividing by the wealth of the richest agent. All the distributions tend to skew to the right as $s$ increases, and therefore the total wealth of the system tends to be more concentrated for few agents. The Log-Pearson distribution constitutes a useful approach to sets of skewed data which have been extensively discussed in the context of hydrology \cite{Singh.1998}. Formally, a random variable $X$ distributes as Log-Pearson if the logarithms of the variable $Y=\ln(X)$ distributes as a generalized gamma distribution, i.e. a three parameter gamma distribution localized at $\zeta \neq 0$. Thus, the density function is defined as follows:

\begin{equation}
f(w)=\frac{1}{\delta w \Gamma(\gamma)}\left(\frac{\ln(w)-\zeta}{\delta}\right)^{\gamma-1} \exp\left(-\frac{\ln(w)-\zeta}{\delta}\right).
\end{equation}

The logarithm transformation of wealth in the density function is specially suitable for the cases of $s\rightarrow \lambda$ where the term $\frac{1}{\lambda - s}$ introduces a certain level of volatility into the model, as we remarked before. Under that transformation the distributions are fitted to three parameter gamma distributions, a generalization of gamma-like distributions of the CC model with a localization parameter. The right-skewness induced by $s$ is studied through the Gini index which is shown in figures \ref{Fig:LP}-a to \ref{Fig:LP}-d. For every case studied we find that the index tends to increase as $s$ become larger, thus the saving $s$ reinforces the wealth inequality between economic agents. Nevertheless for fixed values of $s$ it is found that the distribution becomes more equitable as $\lambda$ becomes larger, as it is shown in figure \ref{Fig:LP}-d. This last result is coherent with the study of the behavior of the Gini index on the CC model made by Ghosh \textit{et al.} \cite{Ghosh.2016}, where it was obtained that the Gini index decreases as values of $\lambda$ are higher.

\section{Economic growth and mean field approach}
According to the dynamics of the model, the total wealth increases as:
\begin{equation}
W(t+1)=W(t) + s\frac{(1-\lambda)[w_i(t)+w_j(t)]}{\lambda-s}.
\label{eq:Wt}
\end{equation}

Additionally, the measure of total production at every transaction is defined by adding the trading prices of goods multiplied by the amount produced of each one:  $Y_w(t)=  p_x X + p_z Z$. Using equations (\ref{eq:QQpx}) and (\ref{eq:QQpz}), the previous expression becomes:
\begin{equation}
Y_w(t)= \frac{(1-\lambda)[w_i(t)+w_j(t)]}{\lambda-s}.
\label{eq:Yt}
\end{equation}

It is clear that both equations depends on the behavior of the individual wealth. Furthermore, both variables are related by the state equation:
\begin{equation}
W(t+1)=W(t) + sY_w(t).
\label{ec:QQWt2}
\end{equation}

Considering a continuous time horizon the last expression is equivalent to $\dot{W}(t)=sY_w(t)$ which constitutes a basic macroeconomic assumption embedded into the frame of the Solow Model \cite{Solow.1956}. A further discussion of this fact within the context of Piketty's studies on wealth inequality is proposed in the next section. For the purpose of this section we limit the analysis to the emergent behavior of individual wealth which drives the total wealth and the total production as it has been pointed out. We proceed in an analogous way of the mean field approach suggested by Bouchaud \textit{et al.} in their model of wealth condensation \cite{Bouchaud.2000}. First, we are interested in the behavior in time of any economic agent $k$. According to the dynamics of the model, the individual wealth of the agent $k$ at any time $t+1$ has a probability $1/N$ of increasing from $w_k(t)$ due to the interaction defined by the equation (\ref{eq:QQdwi}) and a probability  $1/N$ of increasing due to the interaction (\ref{eq:QQdwj}). Hence, assuming that any agent $k$ feels a mean influence from its environment give by the average wealth over all agents $\bar{w}(t)=\sum_{i=1}^{N}w_i(t)$, we redefine both equations as follows:
\begin{equation}
\Delta w_k' = \frac{1-\lambda}{\lambda - s}\left[(s-(1-\varepsilon)\lambda)w_k(t) + \varepsilon\lambda \bar{w}(t) \right],
\label{eq:QQdw'}
\end{equation}
\begin{equation}
\Delta w_k''= \frac{1-\lambda}{\lambda - s} \left[(1-\varepsilon)\lambda \bar{w}(t) + (s - \varepsilon\lambda)w_k(t)\right].
\label{eq:QQdw''}
\end{equation}

Note that the equations (\ref{eq:QQdw'}) and (\ref{eq:QQdw''}) constitute an approximate form of equations (\ref{eq:QQdwi}) and (\ref{eq:QQdwj}) where we assume that the interaction with the other agent can be reduce to an average over all economic agents. Under that approach, the behavior in time of the individual wealth is given by:
\begin{multline}
w_k(t+1)=w_k(t) + \frac{1}{N} \left\{ \Delta w_k' + \Delta w_k'' \right\} \\
=w_k(t) + \frac{(1-\lambda)}{N(\lambda-s)}\left\{ 2s w_k(t) + \lambda \left[ \bar{w}(t) - w_k(t) \right]  \right\}.
\label{eq:wk1_dis}
\end{multline}

Taking the limit $\Delta t \rightarrow 0$ we get the continuous form of the equation (\ref{eq:wk1_dis}):
\begin{equation}
\frac{d w_k(t)}{dt}=gw_k(t) + J\left[ \bar{w}(t) - w_k(t) \right],
\label{eq:wk1_cont}
\end{equation}
where $g=\frac{2s(1-\lambda)}{N(\lambda-s)}$ and $J=\frac{\lambda(1-\lambda)}{N(\lambda-s)}$. The equation (\ref{eq:wk1_cont}) is analogous to the process of evolution of total wealth in the Bouchaud and Mezard model \cite{Bouchaud.2000}, nevertheless the factor $g$ which gives account of the growth of the wealth has not a stochastic behavior related with a Winner process which leads to wealth condensation. Integrating the equation (\ref{eq:wk1_cont}) we find that the average wealth evolves in time as $\bar{w}(t)=\bar{w}_0\exp\left(gt\right)$, where the constant $\bar{w}_0$ is the average wealth per agent at $t=0$ that depends on the initial conditions of the system.

\begin{figure}[!ht]
\centering
\includegraphics[scale=0.125]{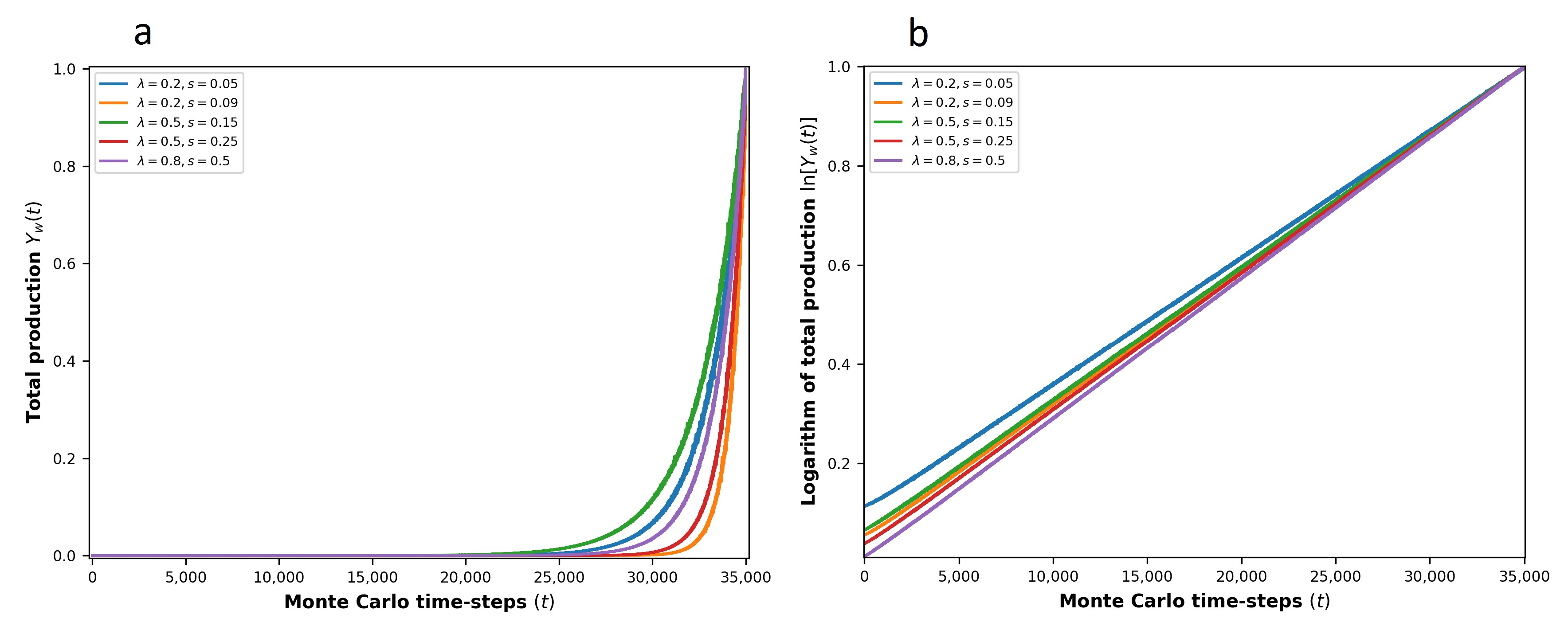}
\caption{\textbf{Economic growth.} (\textbf{a}) The behavior simulated data of production for different cases of $s$ and $\lambda$ is well-fitted by exponential function $Y_w(t)= Y_{w_0} \exp(gt)$, where $g=\frac{2s(1-\lambda)}{N(\lambda-s)}$ and $Y_{w_0}=2\bar{w}_0$. (\textbf{b}) The lineal behavior of semi-log plots ensure the exponential approximation to simulated data. All the results are normalized to $1$ by dividing between the maximum value of production at every interval of time to be presented in the same scale.}
\label{Fig:Yt}
\end{figure}

The behavior of the average wealth in time drives the evolution of production. Hence, under the mean field approximation it grows as $Y_w(t)=Y_{w_0}\exp(gt)$, where $Y_{w_0}=2\bar{w}_0$ and $g$ is the rate of economic growth which has the analytical form discussed above. We tested this result through Monte Carlo simulation method described in the previous section, in the figure \ref{Fig:Yt}-a we show the behavior of production computing directly from the simulation using the relation  $Y_w(t)=  p_x X + p_z Z$ defined by equations (\ref{eq:QQpx}) and (\ref{eq:QQpz}). It is clear from the semi-log plot of production curves shown in the figure \ref{Fig:Yt}-b that the exponential approximation to the evolution of total productions constitutes a good tool to describe its macroscopic behavior. Furthermore, in the table \ref{tb:Results} we compare fitting from simulated data with the analytical relation obtained for $g$ using $N=1000$. For all the cases considered, the relative error is lower than $0.3\%$ which enables to conclude that the mean field theory constitutes a good approach to the model.

\section{Emergence of Piketty's second fundamental \\ law of capitalism}
It is clear from the kinetic dynamic of the model that the income $Y_w(t)$ is exclusively related with production due to return on individual capital. A more accurate approach to economic phenomena should take account of income due to labor. In this line, we suppose that at every time-step every economic agent receives a unity of salary $\bar{Y}_L$ corresponding to income due to labor. Thus, the evolution in time of total income is given by the sum of income due to production and income coming from N unities of average salary $Y(t)=Y_w(t)+N\bar{Y}_l$. This identity is coherent with the macroeconomic framework of Piketty's studies on economic inequality, where the total income of a nation can be broken down in income from capital and income from labor \cite{Piketty.2014,Piketty.2015}. Now, we suppose that at every time step every economic agent saves a rate $s$ of its salary, therefore the individual wealth increases as $w_k(t+1)=w_k(t)+s\bar{Y_l}$. This fact introduces a slight modification into the dynamic of the model, nevertheless the total wealth still is depends on the non-conservative kinetic process in such a way that it increases as $W(t+1) = W(t) + \Delta w_i + \Delta w_j + sNY_l$. Note that the pure kinetic dynamic introduces in the previous section has moved away to an heterogeneous agent-based model.

\begin{figure}[!ht]
\centering
\includegraphics[scale=0.12]{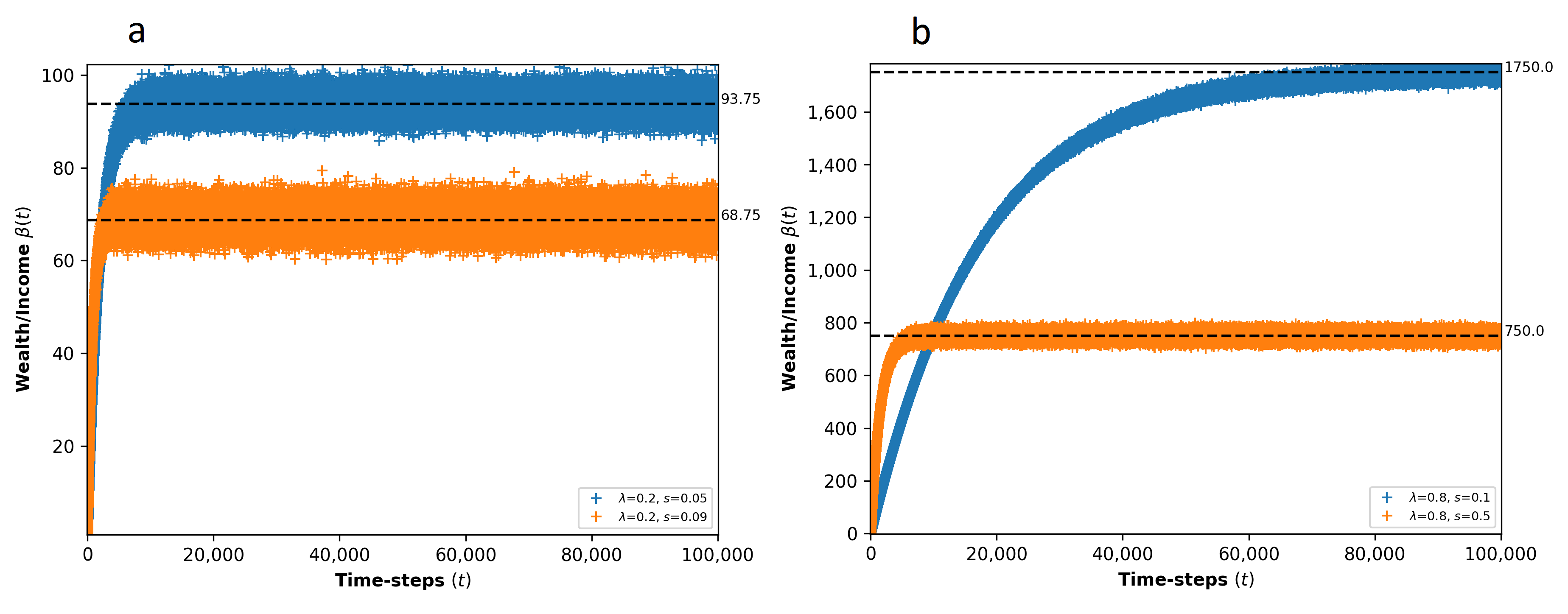}
\caption{\textbf{Piketty's second fundamental law of capitalism.}  On the long run the ratio between total wealth and income $\beta(t)=\frac{W(t)}{Y(t)}$ tends to the ratio $\frac{s}{g}$. This relation is tested for different values of $s$ and $\lambda$ using the simulated data for total wealth and production, and the analytical value of $g$.}
\label{Fig:Bt}
\end{figure}

Under the last modification the model still holds the identity $W(t+1)=W(t)+sY_w(t)+sNY_L$ which leads to $\dot{W}(t)=sY(t)$. At every time-step each economic agent $i$ and $j$ involved in a trading rents over an individual wealth previously increased by $s\bar{Y}_L$ due to income from salary. Thus, the kinetic dynamics induced by the interaction terms $\Delta w_{i,j}$ leads to the an exponential behavior $Y(t)=(Y_{w_0}+Y_L)\exp(gt)$. This result is still consistent with the mean field approach proposed in the previous section where the rate of economic growth  $g$ holds the same relation in terms of exogenous parameters $\lambda$ and $s$. Furthermore the macroscopic emergent behavior of the model, as well as the microscopic dynamics, described till now remain invariant under the sum of the constant term related to the salary.

As we stated before, the macroscopic behavior emergent from the dynamics of the kinetic wealth-exchange model of economic growth satisfies the macroeconomic hypothesis of Solow model \cite{Romer.2006}. The behavior of the total income and the total wealth in time leads to an important emergent property described by Thomas Piketty as the second fundamental law of capitalism. According to the author, in the long-run the ratio between wealth and income $\frac{W(t)}{Y(t)}$ tends to the ratio between the average saving rate of the population, which for the case of the model is fixed by the exogenous parameter $s$, and the rate of economic growth $g$ \cite{Piketty-Zucman.2013}. Specifically, the ratio $W(t)/Y(t)$ is defined within the frame of Piketty's work on economic inequality as the variable $\beta$, and its long-run behavior is induced from the macroeconomic result proposed in the Solow model of economic growth \cite{Piketty-Saez.2014,Solow.1956}. Replacing the function obtained through the mean field approach for the behavior of income in time  $Y(t)=(Y_{w_0}+Y_L)\exp(gt)$ in the equation $\dot{W}(t)=sY(t)$ and integrating the equation we achieve the relation for $W(t)$:
\begin{equation}
W(t)=\frac{s}{g}Y(t) + [W_0 - \frac{s}{g}(Y_{w_0}+N\bar{Y}_l)].
\label{eq:Wt3}
\end{equation}

Now, dividing between $Y(t)$ the relation becomes:
\begin{equation}
\beta(t)=\frac{s}{g} + [W_0 - \frac{s}{g}(Y_{w_0}+N\bar{Y}_l)] \frac{1}{Y(t)},
\label{eq:Bt}
\end{equation}
which gives us an idea of the behavior in time of the ratio $\beta (t)$. It is clear that in the limit $t\rightarrow \infty$ the equation (\ref{eq:Bt}) tends to:
\begin{equation}
\lim_{t\rightarrow \infty}\beta(t)=\frac{s}{g},
\label{eq:limBt}
\end{equation}
due to the exponential behavior of $Y(t)$. This is a mathematical formalization of the second fundamental law of capitalism, where the concept of infinity is introduced in the sense of a very large but reasonable period of time over which the economy reaches a steady state. The last result is a general property of the model which is satisfied too for the case of non income from labor. In this case we find that $W_0 - \frac{s}{g}(Y_{w_0})=0$ and therefore $\beta=\frac{s}{g}$ for every $t$. The behavior of $\beta (t)$ from simulated data is shown in figures \ref{Fig:Bt}. Additionally, in the table \ref{tb:Results}, we show the results from testing the relation $\frac{s}{g}$  using the analytical form of $g$ to predict the limit of simulated data. So, we find that the relative error between both results is less than $0.15\%$ for every considered case.

\section{Discussion and conclusions}

The macroeconomic properties discussed in the last section constitute the core of the extension of the CC model proposed in this paper. As we have shown, the kinetic wealth-exchange model that we present here leads to a macroeconomic scenario coherent with neoclassical theory of economic growth and recent studies on economic inequality. The kinetic dynamics of the model induce a macroscopic phenomenon of economic growth which is driven by the exponential behavior of average individual wealth, which corresponds to the economic growth in the context of the Solow model. This result is obtained analytically from the dynamics of the model via a mean field approach and it is tested via a Monte Carlo simulation. Its consequences over the total wealth and the total production constitute the necessary elements to reproduce the Piketty's second fundamental law of of capitalism \cite{Piketty.2014} as an emergent property of the system.

The microfounded perspective of the the kinetic wealth-exchange model where the notion of aggregation is addressed from the point of view of microscopic dynamics between economic agents leads to important macroeconomic results which contributes to close the gap between microeconomics and macroeconomics through the study of agent-based models. In this line, the model proposes a noteworthy contribution to the study of economics from the point of view of statistical physics.

Regarding the forthcoming horizon of studies opened by this kinetic wealth-exchange model, it is important to highlight the role of the introduction of the labor income factor into the dynamics of the agents in the system. It is clear that for the case of $\bar{Y}_L=0$ the share of capital on the total income is 1, i.e. all the income is due to the return of wealth as consequence of production. Nevertheless, for the case of $\bar{Y}_l \neq 0$ the share of capital income depends on the ratio between the income related to capital and the income coming from labor. This fact suggests a way to tackle forthcoming studies on the macroeconomic theory around Piketty's work on economic inequality through the microfounded framework proposed by this kinetic wealth-exchange model of economic growth.

\newpage

\begin{center}\vspace{1cm}
\scalebox{0.9}{%
\begin{tabular}{|c|c||c|c|c||c|c|c|}
\toprule
\rule{0pt}{15pt} \centering $\lambda$ & $s$ & $g_{sim}$ \scriptsize{$\times 10^{-5}$} & $\frac{2s(1-\lambda)}{N(\lambda - s)}$\scriptsize{$\times 10^{-5}$} & error \scriptsize{$\%$} & $\frac{W(t)}{Y(t)}$ & $\frac{s}{g}$ & error \scriptsize{$\%$} \\
\midrule
0.2 & 0.05 & $53.31$ & $53.33$ & 0.038 & 93.79 & 93.75 & 0.043 \\
0.2 & 0.09 & $130.7$ & $130.9$ & 0.153 & 68.84 & 68.75 & 0.131 \\
0.5 & 0.15 & $42.85$ & $42.86$ & 0.023 & 350.06 & 350.00 & 0.017 \\
0.5 & 0.25 & $99.99$ & $100$   & 0.01 & 250.15 & 250.00 & 0.06 \\
0.8 & 0.1 & $5.714$  & $5.714$ & 0 & 1750.04 & 1750.00 & 0.002 \\
0.8 & 0.5 & $66.64$  & $66.66$ & 0.03 & 750.18 & 750.00 & 0.024 \\
\bottomrule
\end{tabular}}
\captionof{table}{The estimated rate of economic growth from simulated data $g_{sim}$ is well predicted by the relation $g= \frac{2s(1-\lambda)}{N(\lambda - s)}$. This fact is shown for fixed values of $\lambda= \{0.2$, $0.5$, $0.8\}$ and different values of $s$. Additionally, the model satisfies the Piketty's second fundamental law of capitalism which establish that on the long run $\frac{W(t)}{Y(t)}=\frac{s}{g}$. This relation is tested for different values of $s$ and $\lambda$ using the simulated data for total wealth and production, and the analytical value of $g$.}
\label{tb:Results}
\end{center}\vspace{1cm}

\end{document}